# Reconciling Multiple Connectivity Scores for Drug Repurposing

**Kewalin Samart**[1,2,3,*], **Phoebe Tuyishime**[1,3,4,*], **Arjun Krishnan**[3,5,#], **Janani Ravi**[1,#]

[1]Pathobiology and Diagnostic Investigation, Michigan State University; [2]Mathematics, Michigan State University; [3]Computational Mathematics, Science and Engineering; [4]Food Science and Nutrition, Michigan State University; [5]Biochemistry and Molecular Biology, Michigan State University.

*These authors contributed equally to this work.

[#]Corresponding authors: arjun@msu.edu; janani@msu.edu.

## Abstract

The basis of several recent methods for drug repurposing is the key principle that an efficacious drug will reverse the disease molecular 'signature' with minimal side-effects. This principle was defined and popularized by the influential 'connectivity map' study in 2006 regarding reversal relationships between disease- and drug-induced gene expression profiles, quantified by a disease-drug 'connectivity score.' Over the past 15 years, several studies have proposed variations in calculating connectivity scores towards improving accuracy and robustness in light of massive growth in reference drug profiles. However, these variations have been formulated inconsistently using various notations and terminologies even though they are based on a common set of conceptual and statistical ideas. Therefore, we present a systematic reconciliation of multiple disease-drug similarity metrics ($ES$, $css$, $Sum$, $Cosine$, $XSum$, $XCor$, $XSpe$, $XCos$, $EWCos$) and connectivity scores ($CS$, $RGES$, $NCS$, $WCS$, $Tau$, $CSS$, $EMUDRA$) by defining them using consistent notation and terminology. In addition to providing clarity and deeper insights, this coherent definition of connectivity scores and their relationships provides a unified scheme that newer methods can adopt, enabling the computational drug-development community to compare and investigate different approaches easily. To facilitate the continuous and transparent integration of newer methods, this article will be available as a live document (https://jravilab.github.io/connectivity_scores) coupled with a GitHub repository (https://github.com/jravilab/connectivity_scores) that any researcher can build on and push changes to.

## Key points

- Connectivity mapping is a powerful approach for drug repurposing based on finding drugs that reverse the transcriptional signature of a disease, quantified by a connectivity score.
- Though a number of similarity metrics and connectivity scores have been proposed until now, they have been described using inconsistent notations and terminologies to refer to a common set of concepts and ideas.
- Here, we present a coherent definition of multiple connectivity scores using a unified notation and terminology, along with delineating the clear relationship between these scores.
- Our unified scheme can be adopted easily by newer methods and used for systematic comparisons.
- The live document and GitHub repository will allow continuous incorporation of newer methods.

# Introduction

The past few decades has seen a rapid increase in computational, experimental, and clinical drug repositioning/repurposing approaches owing to the appeal of reduced costs and drug discovery time [1–3]. Drug repurposing works on the principle that drugs have multiple modes of action, targets, and off-targets, that can be exploited to identify new indications [1]. This principle has been leveraged to identify novel therapeutic candidates for several diseases [1,4]. Approaches and resources for drug repurposing have been broadly summarized and discussed elsewhere [2,5]. With the accumulation of massive drug and disease data collections, computational methods and databases have now become an indispensable component of the drug repurposing workflow [2,6]. Nearly all these methods leverage high-throughput gene expression profiles abundantly available for drugs and diseases to find novel associations [7–9]. These expression profiles can be used to derive a characteristic molecular imprint, *i.e.,* a signature, of a disease or drug perturbation in a tissue [10]. Large compendia of such transcriptomic signatures have been created for thousands of drugs based on the differential gene expression of various cell lines with or without drug perturbation. Computational methods then use these compendia to predict repurposed candidates for a disease either based on the (dis)similarity of a drug's expression signature to that disease's expression signature [11] or based on similarity to the signatures of other drugs previously linked to the disease [12,13].

In this article, we will focus on these widely-used expression-based methods for drug repurposing collectively referred to as "drug-disease connectivity analysis" [11]. A typical instance of this analysis is presented in **Figure 1** where novel drug indications for a particular disease of interest are identified based on the extent to which the ranked drug-gene signature is a "reversal" of the disease gene signature ([14,15] **Fig. 1**). Connectivity-based drug repurposing has been used to discover drugs in various cancers and non-cancer diseases [3].

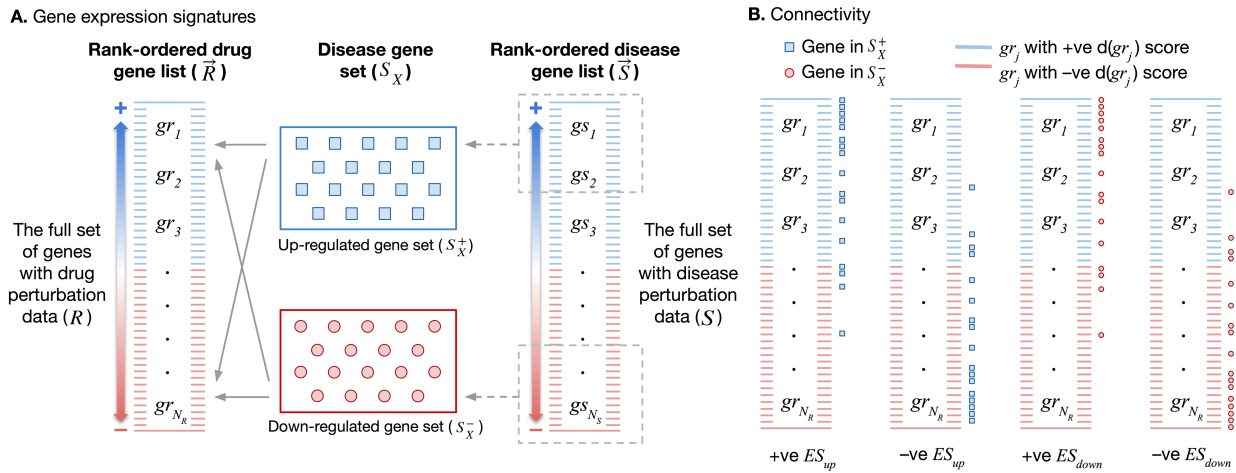

**Figure 1. Drug-disease connectivity. A.** *Gene expression signatures.* Gene expression signatures. $\vec{R}$ and $\vec{S}$ are rank-ordered drug and disease gene expression signatures going from the most significantly up-regulated genes to the most significantly down-regulated genes. $S$ is the full set of genes with disease data. Without any loss in generality, only the subset of disease genes that are also part of $R$ are considered throughout (*i.e.,* $S \subseteq R$ ). $S^+$ and $S^-$ correspond to the set of most up- and down-regulated sets of disease genes, respectively. **B.** *Connectivity*. Positions of $S^+$ and $S^-$ disease genes in the ranked drug list, $R$, determine the signs of enrichment scores ($ES$; $ES_{up}$, $ES_{down}$). Positive connectivity is defined as the case when the disease signature and drug profile show similar perturbations, *i.e.,* when $ES_{up}$ is positive and/or when $ES_{down}$ is negative. This happens when $S^+$ predominantly appears towards the top of the drug profile or when $S^-$ appears predominantly towards the bottom of the drug profile (scenarios 1 and 4). Conversely, negative connectivity is defined as the case when the disease signature and drug profile show dissimilar perturbations, *i.e.,* when $ES_{up}$ is negative and/or when $ES_{down}$ is positive. This happens when $S^+$ predominantly appears towards the bottom of the drug profile or when $S^-$ appears towards the top of the drug profile (scenarios 2 and 3). Negative connectivity indicates drug reversal of disease signature.

From its inception in 2006, the exact method for connectivity analysis has evolved, with a series of proposed modifications over the past decade and a half (**Fig. 2**). The first method for connectivity analysis [7] builds on the classic paper by Subramanian *et al.,* 2005 [16] that proposed the Gene Set Enrichment Analysis (GSEA) method. GSEA uses a modified Kolmogorov-Smirnov statistic (KS) [17] – referred to as "enrichment score" ($ES$) – to evaluate if genes in a certain pathway appear towards the top or bottom of a gene (differential) expression profile. Lamb *et al.,* 2006 [7] built a reference database (Connectivity Map or CMap, which we refer to as CMap 1.0 in this article) with gene expression profiles for thousands of small molecules and proposed the first method for connectivity analysis based on GSEA. This method compares a query signature (disease) to each of the ranked drug-gene expression profiles in their reference database and ranks all the drugs based on their connectivity scores. A connectivity score ranges between -1 (indicating a complete 'drug-disease' reversal) and +1 (indicating perfect 'drug-disease' similarity). Another study adapted this connectivity score calculation and used it to find compounds in the L1000 LINCS collection [8] that could be repurposed for three cancer types [18]. This study quantified the reversal relationship between the drug and disease by computing the Reverse Gene Expression Signature ($RGES$). Finally, CMap 1.0 itself was further updated by expanding the Library of Integrated Network-based Cellular Signatures (LINCS) L1000 to more than 1.3 million profiles [19] (referred to as CMap 2.0 in this article). Along with the expansion of data, the CMap 2.0 study also proposed another variation of the connectivity score called the weighted connectivity score that uses GSEA's weighted Kolmogorov-Smirnov enrichment statistic along with ways to normalize the resulting score and correcting them further to account for background associations.

Another class of connectivity scores has been developed that uses the level of differential expression of genes in its calculations, thus distinguishing itself from the approaches mentioned above that invariably use just the gene ranking [20–23]. Jointly referred to as pairwise similarity measures, they use the drug/disease differential-expression values of either all genes or just the most perturbed genes (called 'extreme' metrics). One such score called connection strength score ($CSS$) reflects the strength of the correlation between the signed ranks of genes in the disease and drug profiles [23]. In other cases, final scores are derived by summing gene scores ($Sum$, $XSum$) or by calculating the correlation between the drug and disease profiles using any one of several correlation metrics ($XSpe$, $XCor$, $Cosine$, $XCos$) [20,21]. The cosine metrics have been further modified to reduce the impact of lowly-expressed genes ($EWCos$) [22]. With the advent of numerous connectivity scores, a recent study has developed an approach called the Ensemble of Multiple Drug Repositioning Approaches ($EMUDRA$) that normalizes and integrates four metrics ($EWCos$, $Cosine$, $XSpe$, and $XCor$) into one score [22].

Connectivity scores and methodologies have been evaluated in the past to assess their performance in predicting drug-drug relationships or drug-disease relationships. The performance of CMap 1.0 was evaluated in predicting drug-drug relationships using the Anatomical Therapeutic Chemical classification [20,24], and in predicting drug-disease relationships [25]. Furthermore, a recent review [26] assessed advances that have been made in CMap 1.0 and computational tools that have been applied in the drug repurposing and discovery fields. Lin *et al.,* 2019 [27] further evaluated connectivity approaches that use L1000 data [8], including six different scores that are used to predict drug-drug relationships.

All these proposed variations of the connectivity score share a common set of conceptual and statistical ideas. Yet, they have been formulated inconsistently using varied notations and terminologies in the original papers and in the aforementioned evaluation studies. This lack of consistency in the precise formulaic notation makes it difficult to seamlessly understand the subtle differences and the intuition underlying each score. For example, the connectivity score referred to as Reverse Gene Expression Score, "$RGES$" [18], directly builds on the Connectivity Score, "$CS$" [7]. Another example is the Weighted Connectivity Score, "$WCS$" in CMap 2.0 [19] that is a bi-directional weighted version of "$ES$" used in GSEA [16]; in this case, they are named and notated quite differently though they are essentially direct, simple variants of each other. "$ZhangScore$" in [27] and "$WSS$" in [22] refer to the connection strength $C$ in [23]. In this article, we develop a systematic scheme that defines in the aforementioned methodologies using consistent notations and terms. Additionally, we provide summary tables throughout the article to relate our consistent scheme with the previously published ones.

# A taxonomy of connectivity scores

We begin by creating a standardized set of notations and terms to denote the various concepts and quantities required to define the different connectivity scores. In its most widely-used form, a connectivity score between a disease and a drug is computed by comparing the genes significantly up- ($S_X^+$) and down-regulated ($S_X^-$) by the disease (relative to a healthy control) to a ranked list of genes ordered based on their differential expression in response to a drug ($\vec{R}$). A good connectivity score usually manifests as a lower negative value since it is designed to indicate a reversal relationship between the disease and the drug on genes. Such a score is achieved when genes in $S_X^+$ appear at the bottom of $\vec{R}$ and/or when genes in $S_X^-$ appear at the top of $\vec{R}$. When there is no relationship or when $S_X^+$ appears at the top and/or when $S_X^-$ appears at the bottom of $\vec{R}$ (indicating a similarity between the disease and drug signatures), the drug is considered unlikely to be efficacious in treating that disease. These scenarios are depicted in **Figure 1**. The general notations, which we use throughout this work, are presented in **Table 1, Figure 2**.

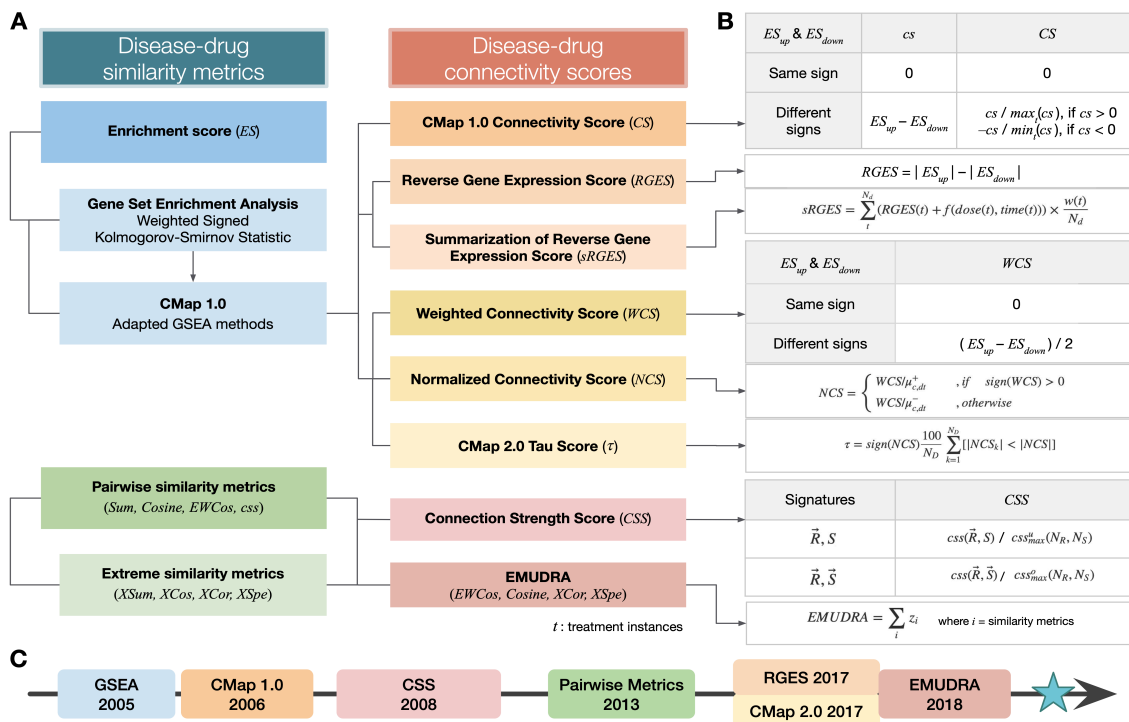

**Figure 2. A taxonomy of connectivity scores. A.** Relationship between disease-drug similarity metrics and connectivity scores. **B.** Detailed definitions of connectivity scores in **A**. **C.** A brief history of connectivity scores. CMap-based connectivity scores: **2005–06.** The first connectivity score, $CS$, was proposed in CMap 1.0 [7]. This score was derived based on a modified KS statistic proposed as part of GSEA [16]. **2017.** a) A later study [18] proposed a new score $RGES$ that combines the enrichment scores (used by $CS$) in a new way and shows an inverse correlation with drug efficacy. The same study also proposed heuristics to combine the $RGES$ values across multiple instances of the same drug, derived from different dosages and treatment times, into a summarized score $sRGES$. b) The CMap 2.0 study [19], which included the generation of the massive LINCS resource, proposed yet another set of connectivity scores, $WCS$, $NCS$, and $\tau$, that build on $CS$. While $CS$ is based on gene ranks alone, $WCS$ uses a "weighted" enrichment score from GSEA [16] that takes the gene perturbation levels into account. $WCS$ scaled by appropriate mean values gives $NCS$. Finally, the entire LINCS dataset is exploited to perform an additional permutation-based correction of $NCS$ to finally obtain $\tau$. Pairwise similarity metrics and connectivity scores: **2008.** The connectivity score $CSS$ was proposed to include the ranks of all the genes along with their direction of perturbation to calculate drug-disease similarity, followed by a correction based on scores for all other drugs in the database [23]. **2013–2018.** Others [20–22] proposed a set of simple pairwise similarity metrics to calculate drug-disease associations that incorporate the magnitude and direction of gene differential expression under both the drug and the disease (more below). A new connectivity score, $EMUDRA$, integrates multiple pairwise scores to leverage the benefits of all of them [22].

Building on these general notations and terms, in the rest of this article, we develop and present a systematic scheme that defines the formulations of several drug-disease similarity metrics and connectivity scores using consistent notations and terms (**Table 1**; **Fig. 2**), detailed formulation, and summary tables (**Tables 2–8**; **Fig. 3**) that will enable researchers to relate our consistent scheme back to the notations and terminology used in the original publications.

**Table 1. General Notations**

| Notation | Description |
|---|---|
| $R$ | the full set of genes with drug perturbation data |
| $S$ | the full set of genes with disease perturbation data (*i.e.,* query) (**Fig. 1**); Without any loss in generality, only the subset of disease genes that are also part of $R$ are considered throughout (*i.e.,* $S \subseteq R$ ). |
| $gr_i, gs_i$ | $i^{th}$ gene in set $R$ or set $S$ (*i.e.,* drug or disease gene) |
| $N_R, N_S$ | number of genes in gene sets $R$ or $S$ |
| $S^+, S^-$ | disease up- or down-regulated genes; $S^+ \subseteq S$, $S^- \subseteq S$, $S^+ \cup S^- = S$ |
| $R_X, S_X$ | subset of drug genes ($R$) or disease genes ($S$) with the most extreme gene scores (either from the top or bottom) defined based on a user-specified threshold of fold-change and/or significance; $R_X \subseteq R$; $S_X \subseteq S$ |
| $S_X^+, S_X^-$ | the up- and down-regulated subsets of SX, the genes with the extreme disease gene scores; $S_X^+ \cup S_X^- = S_X$ |
| $\vec{R}, \vec{S}$ | rank-ordered drug or disease gene list (*i.e.,* ordered version of $R$ or $S$) from the highest to the lowest gene scores (*e.g.,* Figure 1 shows $\vec{R}$) |
| $\overrightarrow{R_{abs}}, \overrightarrow{S_{abs}}$ | absolute rank-ordered drug or disease gene list from the highest to the lowest absolute gene scores |
| $\overrightarrow{R_X}, \overrightarrow{S_X}$ | rank-ordered gene list for $R_X$ and $S_X$ |
| $r_{drg}(), r_{dis}()$ | rank function for drug or disease that takes one or more genes as input and returns a vector of their ranks in $\vec{R}$ or $\vec{S}$, respectively |
| $r_{drg}^{abs}(), r_{dis}^{abs}()$ | absolute rank function for drug or disease that takes one or more genes as input and returns a vector of their absolute ranks in $\vec{R}$ or $\vec{S}$, respectively |
| $v_{drg}(), v_{dis}()$ | score function for drug or disease that takes one or more genes as input and returns a vector of their gene scores in $\vec{R}$ or $\vec{S}$, respectively |
| $sgn_{drg}(), sgn_{dis}$ | sign function for drug or disease that takes one or more genes as input and returns the signs of their gene scores (+1 or –1) in $\vec{R}$ or $\vec{S}$, respectively |
| $t$ | each treatment instance (*i.e.,* a treated-and-vehicle-control pair) that results in a single drug profile $R$ or $\vec{R}$. |
| $N_D$ | total number of drug profiles ($R$ or $\vec{R}$) in the reference database |
| $N_d$ | number of drug profiles ($R$ or $\vec{R}$) in the reference database that corresponds to a specific drug $d$ |

## Gene Set Enrichment Analysis (GSEA)

Nearly all connectivity scores developed thus far begin with the calculation of some form of an Enrichment Score ($ES$) that captures the relationship between a drug and a disease. The basis of all these $ES$ formulations is the Gene Set Enrichment Analysis (GSEA) [16], which was originally developed to assess the enrichment (over-representation) of predefined biological gene sets (*e.g.,* pathways, targets of a regulator) at the top or bottom of a list of genes ranked by their extent of differential expression in response to an experimental factor of interest. Enriched gene sets are then hypothesized to be biologically relevant to that experimental factor. When adapted to the question of drug repurposing, a method like GSEA can be used to assess the enrichment of sets of genes associated with a disease at the top or bottom of a list of genes ranked by their extent of differential expression in response to a drug (**Fig. 1**). In this section, we present the formulation of $ES$ using our new, consistent notation (**Table 2**; **Fig. 3**).

### *Enrichment Score (ES)*

GSEA is a weighted signed version of the classical Kolmogorov-Smirnov test. It takes two inputs: i) a disease gene set composed of a set of genes significantly perturbed in response to a disease (denoted $S_X \subseteq S$), and ii) a rank-ordered list ($\vec{R}$) of drug genes (in decreasing order of $v_{drg}(gr_j)$, a score based on the level differential-expression of each gene $gr_j$ in response to the drug). Using these two inputs, GSEA quantifies the level of association between the disease and the drug by calculating an enrichment score ($ES$) based on the following steps:

1. For each position $i$ in the rank-ordered list ($\vec{R}$) from top to bottom,

    1.1. if the gene is in $S_X$, calculate:

    $$P_{hit}(S_X, i) = \sum_{\substack{gr_j \in S_X \\ j \leq i}} \frac{|v_{drg}(gr_j)|^{w_{ES}}}{N'_{S_X}}, \quad \text{where} \quad N'_{S_X} = \sum_{gr_j \in S_X} |v_{drg}(gr_j)|^{w_{ES}}$$

    1.2. if the gene is not in $S_X$, calculate:

    $$P_{miss}(S_X, i) = \sum_{\substack{gr_j \notin S_X \\ j \leq i}} \frac{1}{N_R - N_{S_X}}$$

    1.3. calculate the positional enrichment score ($es_i$)

    $$es_i = P_{hit}(S_X, i) - P_{miss}(S_X, i)$$

2. Finally, calculate the final enrichment score ($ES$):

$$ES = max_i(es_i), \qquad (1.1)$$

the maximum positional enrichment score. Here, $N_R$ and $N_{S_X}$ are the number of genes in the drug ($R$) and disease ($S_X$) gene sets. $w_{ES}$ is the weight assigned to each position in the drug profile $\vec{R}$. When $w_{ES} = 0$, $N'_{S_X} = \sum_{gr_j \in S} |v_{drg}(gr_j)|^0 = N_S$, which results in

$$P_{hit}(S_X, i) = \sum_{\substack{gr_j \in S_X \\ j \leq i}} \frac{|v_{drg}(gr_j)|^0}{N'_{S_X}} = \sum_{\substack{gr_j \in S_X \\ j \leq i}} \frac{1}{N'_{S_X}}.$$

Thus, $P_{hit}(S_X, i)$ and $P_{miss}(S_X, i)$ are both empirical distribution functions of the positions of the disease genes (*i.e.,* $S_X$) and the positions of the non-disease genes (*i.e.,* $R - S_X$), respectively, in the drug gene list

(1.2)

$\vec{R}$. Therefore, when $w_{ES} = 0$, $ES$ (the signed maximum distance between the two functions) reduces to a signed two-sample Kolmogorov-Smirnov (KS) statistic:

$$ES = max(P_{hit}(S_X, i) - P_{miss}(S_X, i)) = sign(P_{hit}(S_X, i) - P_{miss}(S_X, i)) \times KS$$

where

$$KS = max|F_{S_X}(i) - F_{R-S_X}(i)|$$

is the classical two-sample KS statistic, with $F_{S_X}$ and $F_{R-S_X}$ being the empirical distribution function of $S_X$ and $R - S_X$, respectively, defined as follows:

$$F_{S_X}(i) = \frac{1}{N_{S_X}} \sum_{\substack{j=1 \\ gr_j \in S_X}}^{N_{S_X}} 1_{j \le i}, \qquad F_{R-S_X}(i) = \frac{1}{N_R - N_{S_X}} \sum_{\substack{j=1 \\ gr_j \notin S_X}}^{N_R} 1_{j \le i}$$

where $1_{j \le i}$ is the indicator variable that takes the value 1 whenever $j \le i$ and 0 otherwise.

When $w_{ES} = 1$, $ES$ becomes a weighted signed two-sample KS statistic with each position $j$ in the drug gene list $\vec{R}$ weighted by the drug-response score $v_{drg}(gr_j)$. Setting $w_{ES}$ to one is recommended for GSEA. We point the reader to the original GSEA publication for a discussion of statistics when $w_{ES}$ is set to lesser or greater than one.

**Summary**

- Enrichment score, $ES$, ranges from –1 to +1.
- $ES$ is the maximum deviation from zero encountered between the empirical distributions of the disease and non-disease genes in drug gene list $\vec{R}$.
- A positive $ES$ indicates disease gene set enrichment towards the top of drug gene list $\vec{R}$.
- A negative $ES$ indicates disease enrichment at the bottom of $\vec{R}$.
- When $S_X$ is randomly distributed in $\vec{R}$, the magnitude of $ES$ is small but if a large proportion of genes in $S_X$ is concentrated at the top or bottom of $\vec{R}$, the magnitude of $ES$ is large.
- When calculated separately for genes up- ($S_X^+$) and down-regulated ($S_X^-$) by the disease, good drug candidates that show a reversal relationship with the disease profile have a negative $ES_{up}$ and a positive $ES_{down}$ (**Table 2**; **Fig. 3**).
- Revised notations used in this GSEA section are summarized in **Table 2**.

**Table 2. GSEA Notations**

| Current Notation | Previous Notation | Description |
| --- | --- | --- |
| KS | – | Kolmogorov-Smirnov |
| ES | – | enrichment score |
| $ES_{up}$, $ES_{down}$ | – | $ES$ for up-regulated gene set ($S_X^+$) or down-regulated gene sets ($S_X^-$) |
| $w_{ES}$ | $p$ | the weight assigned to positions in $\vec{R}$ when calculating $ES$ |
| $gr_j$ | $g_j$ | a gene in $\vec{R}$ at index $j$ |
| $v_{drg}(gr_j)$ | $r_j$ | the drug-response score of gene $gr_j$ in drug gene list $\vec{R}$; this score is used to rank the genes in $\vec{R}$ |
| $N_{S'_X}$ | $N_R$ | the sum of absolute drug gene score ($v_{drg}(gr_j)$) of every $\vec{R}$ gene in $S_X$ weighted by $w_{ES}$ |
| $P_{hit}(S_X, i)$ | – | the fraction of genes in $S_X$ ("hits") weighted by their drug gene score ($v_{drg}(gr_j)$) |
| $P_{miss}(S_X, i)$ | – | the fraction of genes not in $S_X$ ("misses") |
| $N_R, N_{S_X}$ | $N, N_H$ | number of genes in $\vec{R}$ or $S_X$ |

## Connectivity Map 1.0: Disease-Drug Connectivity Scores (CMap 1.0)

The connectivity map 1.0 (CMap 1.0) project pioneered the identification of drug candidates based on their ability to reverse disease gene expression profiles [7]. Key to this project was the creation of a large collection of reference gene expression profiles of multiple human cell lines that are treated with 164 small molecules, including approved drugs. The expression profiles were generated using Affymetrix microarrays. The original CMap 1.0 study and several others focused on cancer [28], inflammatory bowel disease [14] and spinal muscular atrophy [29] have used this reference library of drug profiles for drug repurposing. In all these cases, the starting point is a disease "signature" defined by the sets of genes up- and down-regulated in the disease. This signature is compared to each drug profile in the reference library using a GSEA-like analysis that results in an enrichment score ($ES$) for each of the up- and down-regulated disease gene sets separately. The $ES$ captures the level and direction of association of the disease gene set with that drug. Then, the 'up' and 'down' $ES$ are combined into a single connectivity score ($CS$) for the disease with respect to that drug. Finally, for the given disease, drug candidates are identified as those that have low negative $CS$. In this section, we present the formulation of CMap 1.0 using our new, consistent notation (**Table 3**).

### ES Calculation

The drug-disease enrichment score ($ES$) in CMap 1.0 is adapted from GSEA. Instead of using GSEA's signed two-sample KS test formulation that compares the positions of $S_X$ genes to those of $R - S_X$ genes, CMap 1.0 uses a signed one-sample KS test to compare the empirical distribution of the positions of genes in $\vec{R}$ compared to a reference uniform distribution (of disease genes in the drug gene list).

$$ES = \begin{cases} a & , if \quad a > b \\ -b & , if \quad b > a \end{cases} \quad (2.1)$$

where

$$a = \max_{i=1}^{N_S}\left[\frac{i}{N_S} - \frac{r_{drg}(gs_i)}{N_R}\right]$$

$$b = \max_{i=1}^{N_S}\left[\frac{r_{drg}(gs_i)}{N_R} - \frac{(i-1)}{N_S}\right]$$

This formulation is used to calculate an $ES_{up}$ and an $ES_{down}$ value for the genes up- ($S_X^+$) and down-regulated ($S_X^-$) by the disease, respectively.

### Connectivity Score (CS) Calculation — Normalization across treatment instances

These two scores are then used to calculate a raw connectivity score $cs$:

$$cs = \begin{cases} ES_{up} - ES_{down} & , if \quad sign(ES_{up}) \neq sign(ES_{down}) \\ 0 & , otherwise \end{cases}$$

The final connectivity score is calculated by normalizing the raw score by dividing by the maximum or minimum of raw scores across treatment instances, depending on the sign of $cs$, bringing it back to range between –1 and +1:

$$CS = \begin{cases} \dfrac{cs}{max_t(cs)} & , if \quad cs > 0 \\ \dfrac{-cs}{min_t(cs)} & , if \quad cs < 0 \end{cases} \tag{2.2}$$

**Summary**

- $ES_{up}$ and $ES_{down}$ represent the association between the up- ($S_X^+$) and down-regulated ($S_X^-$) disease genes ($S_X$) with the ranked drug gene list ($\vec{R}$).
- $CS$ is the connectivity score that combines $ES_{up}$ and $ES_{down}$ per drug treatment and normalizes them across treatments. Similar to $ES$, $CS$ ranges from –1 to +1.
- Lower $CS$ indicates a better reversal relationship between the disease and the drug.
- Revised notations used in this CMap 1.0 section are summarized in **Table 3**.

**Table 3. CMap 1.0 Notations**

| Current Notation | Previous Notation | Description |
|---|---|---|
| $CS$ | $S^i$ | connectivity score; normalized connectivity score across all treatment instances |
| $t$ | $i$ | treatment instances |
| $cs$ | $s^i$ | connectivity score for each treatment instance |
| $ES$ | $ks$ | enrichment score |
| $r_{drg}(gs_i)$ | $V(j)$ | position of $gs_i$ in $\vec{R}$ |
| $N_R, N_{S_X}$ | $t, n$ | number of genes in $\vec{R}$ and $S_X$ |

## Reverse Gene Expression Scores (RGES)

The Connectivity Map project was subsequently expanded into the NIH library of integrated network-based cellular signatures (LINCS) program by using a cost-effective gene-expression assay called L1000 [19]. The L1000 platform measures only about 1,000 carefully-chosen genes with the rest of the transcriptome estimated by an imputation model trained using publicly available genome-scale expression data [9]. The pilot phase of the LINCS program included data for about 20,000 compounds assayed on about 50 human cell lines across a range of doses to result in over one million L1000 profiles.

The focus of the study by Chen *et al.,* 2017 [18] was to use this LINCS data to not only capture expression-based drug-disease reversal relationships but also evaluate if these reversals correlate with independently-measured drug efficacies. Towards this goal, the authors selected compounds with both efficacy data in ChEMBL [30] and gene expression LINCS data. Using these two datasets, this study showed that the distribution of connectivity scores ($CS$) from CMap 1.0 [7] are enriched at 0 and that these scores do not correlate well with $IC_{50}$ values. To address this issue, the authors proposed a new connectivity score called the Reverse Gene Expression Score ($RGES$). In this section, we present the formulation of $RGES$ using our new, consistent notation (**Table 4**).

In CMap 1.0, the connectivity score for a drug is set to zero if $ES_{up}$ and $ES_{down}$, the enrichment scores for the up- and down-regulated disease gene sets have the same signs. $RGES$, on the other hand, is computed as the difference between absolute values of the two $ES$ values:

$$RGES = |ES_{up}| - |ES_{down}| \tag{3.1}$$

**Summary**

- The $RGES$ connectivity score is based on the difference between the absolute values of the scores of the up- and down-regulated disease genes regardless of whether they are enriched at the top or the bottom of the drug gene list, $\vec{R}$.
- Similar to $ES$ and $CS$, $RGES$ ranges from –1 to +1.
- $RGES$ is inversely correlated with drug efficacy.
- Revised notations used in this $RGES$ subsection are summarized in **Table 4**.

### Summarization of Reverse Gene Expression Score

Since the LINCS dataset contains multiple profiles corresponding to the same drug assayed on multiple cell lines, concentrations, and time points, the study also proposed summarizing a drug's $RGES$ values across these various conditions into a single score called the Summarization of Reverse Gene Expression Score ($sRGES$). $sRGES$ is estimated by first setting the condition that corresponds to 10 μM and 24 hours (the most common in the LINCS database) as the 'reference' condition and setting all other conditions as 'target' conditions. Then, for a specific cell line, a drug's $RGES$ in a target condition is assumed to be dependent on the target condition's dose and time relative to the reference condition, quantified using a heuristic "awarding function" ($f$):

$$f(dose(t), time(t)) = \begin{cases} \alpha, & dose(t) < 10\mu M \text{ and } time(t) < 24 \text{ hours} \\ \beta, & dose(t) < 10\mu M \text{ and } time(t) \geq 24 \text{ hours} \\ \gamma, & dose(t) \geq 10\mu M \text{ and } time(t) < 24 \text{ hours} \\ 0, & dose(t) \geq 10\mu M \text{ and } time(t) \geq 24 \text{ hours} \end{cases}$$

Target conditions are first divided into four groups (as in the equation above), and the value of the function for each target group (*e.g.*, $dose(t) < 10\mu M$ and $time(t) < 24$ hours) is estimated by averaging the difference in $RGES$ between the target group and reference group across all the drugs in the reference database that were profiled in the same cell line in that target condition and the reference condition.

Then, to combine $RGES$ values across cell lines, a weight $w(t)$ is calculated for each treatment that reflects how much that treatment's corresponding cell line, $cell(t)$ is similar to the disease under study:

$$w(t) = \frac{cor(cell(t), disease)}{max_k(cor(cell(k), disease))}$$

Here, the correlation between cell line, $cell(t)$, and the disease, $cor(cell(t), disease)$, is the average of the Spearman correlations between the expression profiles of the cell line and disease of interest, normalized by the maximum correlation between all cell lines and the disease. Finally, $sRGES$ is defined as the following:

$$sRGES = \sum_{t}^{N_d} (RGES(t) + f(dose(t), time(t))) \times \frac{w(t)}{N_d} \tag{3.2}$$

This study shows that these new formulations of the connectivity scores, $RGES$ and $sRGES$, show a correlation with drug $IC_{50}$ values, with drugs with low negative $RGES$ or $sRGES$ tending to have low $IC_{50}$ values.

**Summary**

- The $sRGES$ connectivity score is designed to combine the $RGES$ values based on the difference between the absolute values of the scores of the up- and down-regulated disease genes regardless of whether they are enriched at the top or the bottom of the drug gene list, $\vec{R}$.
- Similar to $ES$ and $CS$, $sRGES$ ranges from –1 to +1.
- $sRGES$ is inversely correlated with drug efficacy.
- Revised notations used in this $sRGES$ subsection are summarized in **Table 4**.

**Table 4. $RGES$ and $sRGES$ Notations**

| Current Notation | Previous Notation | Description |
|---|---|---|
| $RGES$ | – | reverse gene expression score |
| $sRGES$ | – | summarized reverse gene expression score |
| $f(dose(t), time(t))$ | $f(dose(i), time(i))$ | the difference in $RGES$ between a target condition and reference condition, modeled as a function of dose and time |
| $cor(cell(t), disease)$ | $cor(cell(i), disease)$ | the average Spearman correlation between the expression profiles of a cell line $cell(t)$ and the disease of interest |
| $ES$ | $KS$ | enrichment score |
| $N_d$ | $N$ | number of treatments for a given drug ($d$) |
| $t$ | $i$ | treatment instances |

## CMap 2.0 Connectivity Scores

CMap 2.0 is a massive expansion of the L1000 dataset to ~1.4 million profiles, which represent 42K genetic and small molecules perturbed across multiple cell lines [19]. As part of the release of this data, the study also proposed new connectivity score calculations (Weighted Connectivity Score, Normalized Connectivity Score, and Tau Score). Similar to other scenarios outlined above, the CMap 2.0 methodology works by comparing the disease gene set ($S$) (containing the up- ($S^+$) and down-regulated ($S^-$) genes) to reference drug profiles in the L1000 database to get a rank-ordered list of all drugs based on a slightly new formulation of the connectivity score, along with new proposals for normalizing the scores across cell lines and drug types and for correcting the resulting normalized score against the background of the entire reference library. In this section, we present the formulation of CMap 2.0 using our new, consistent notation (**Table 5**).

### Weighted Connectivity Score (WCS)

The disease-drug enrichment score ($ES$) in CMap 2.0 is based directly on GSEA's weighted signed two-sample KS statistic that compares the positions of $S_X$ genes to those of $R - S_X$ genes with the weight $w_{ES}$ set to 1. $ES$ is then used to calculate a Weighted Connectivity Score ($WCS$) that represents a non-parametric disease-drug similarity measure. $WCS$ is defined as follow:

$$WCS = \begin{cases} (ES_{up} - ES_{down})/2 & , if \quad sign(ES_{up}) \neq sign(ES_{down}) \\ 0 & , otherwise \end{cases} \quad (4.1)$$

**Summary**

- The disease-drug similarities ($ES_{up}$ & $ES_{down}$) are computed using the two-sided weighted KS statistic.
- $WCS$ ranges from –1 to +1.
- A positive (or negative) $WCS$ indicates that $S_X$ and $\vec{R}$ are positively (or negatively) related (similar/dissimilar).
- A zero $WCS$ indicates that $S_X$ and $\vec{R}$ are unrelated.
- Revised notations used in this $WCS$ subsection are summarized in **Table 5**.

### Normalized Connectivity Score (NCS)

The Normalized Connectivity Score (*NCS*) was developed to enable the comparison of *WCS* across cell lines and drug type. Given the *WCS* for a disease in relation to a specific drug of a type *dt*, tested in cell line *c*, the corresponding *NCS* is computed by mean-scaling *WCS*:

$$NCS = \begin{cases} WCS/\mu^+_{c,dt} & , if \quad sign(WCS) > 0 \\ WCS/\mu^-_{c,dt} & , otherwise \end{cases} \quad (4.2)$$

Here, $\mu^+_{c,dt}$ and $\mu^-_{c,dt}$ are absolute values of the means of the positive and negative $WCS$ values, respectively. This procedure is identical to that used in the original GSEA for normalizing $ES$ scores to make them comparable across gene sets of different sizes.

### Tau scores

Finally, the Normalized Connectivity Score $NCS$ for a disease to a specific drug (*i.e.,* the $NCS$ for a given disease-drug pair) is converted to a tau ($\tau$) score by comparing it to $NCS$ values of that disease to all the drugs in the reference database (referred to as "touchstone" in CMap 2.0) of the same type $dt$ tested in the same cell line $c$, expressed as signed percentage value between –100 and +100:

$$\tau = sign(NCS) \frac{100}{N_D} \sum_{k=1}^{N_D} [\,|NCS_k| < |NCS|\,] \quad (4.3)$$

Thus, a $\tau$ of 95 indicates that only 5% of drugs in the reference database of the same type and tested in the same cell line (containing $N_D$ drugs) showed stronger connectivity to the disease than the drug of interest. Since any disease is queried against the same fixed drug reference database, $\tau$ values are comparable across diseases.

Another way to calculate a $\tau$ score corresponding to the $NCS$ value for a disease-drug pair is to compare to the $NCS$ values of that specific drug to all the perturbation signatures in a reference database. This comparison will yield a $\tau$ that represents the signed percentage of reference signatures that are less connected to the drug than the disease of interest. In other words, based on this comparison, a $\tau$ of 95 indicates that only 5% of signatures in a reference database showed stronger connectivity to the drug than the disease of interest. Similarly, $\tau$ values in this new setting are comparable across drugs in the reference database.

**Summary**

- The normalized connectivity score $NCS$ was developed to enable the comparison of $WCS$ across cell lines and drug type.
- The tau score ($\tau$) measures further corrects for non-specific associations by expressing the $NCS$ of a given disease-drug pair in terms of the fraction of signatures/profiles in a reference database that exceed this $NCS$ value.
- Tau ($\tau$) ranges from –100 to +100 and a lower negative score reveals a better disease-drug reversal relationship.
- Good tau scores ($\tau$) should range between –95 and –100. A $\tau$ of 95 indicates that only 5% of reference signatures/profiles in the reference database showed stronger connectivity.
- Revised notations used in the $NCS$ and $\tau$ subsections are summarized in **Table 5**.

**Table 5. CMap 2.0 Notations**

| Current Notation | Previous Notation | Description |
|---|---|---|
| $WCS$ | $WTCS$; $w_{c,t}$ | weighted connectivity score; also used to refer to a specific instance of the weighted connectivity score of a given cell line $c$ and perturbagen type $dt$ |
| $c$ | — | cell line |
| $dt$ | $t$ | drug type |
| $k$ | $i$ | index of each drug in the reference database; $k = 1,2,3,…,N_d$ |
| $\mu^+_{c,dt}, \mu^-_{c,dt}$ | $\mu^+_{c,t}, \mu^-_{c,t}$ | absolute values of means of positive and negative raw weighted connectivity scores, respectively |
| $N_D$ | $N$ | total number of drug profiles ($\vec{R}$) in the reference database |
| $S_X$ | $q$ | disease gene set (*i.e.,* query) |
| $\vec{R}$ | $r$ | rank-ordered gene list (drug) |

**Pairwise similarity measures**: All the connectivity scores described above use the enrichment score ($ES$) as the similarity metric, which is a weighted signed two-sample or one-sample KS statistic. However, only the $ES$ used in CMap 2.0 ($WCS$, $NCS$, and $\tau$) incorporates drug gene perturbation values (by setting the weight $w_{ES}$ to $\geq 1$). The $ES$ used in the other scores ($CS$, $RGES$, and $sRGES$) is just based on gene ranks, thereby likely missing several potential drug candidates. Additionally, all these scores (including CMap 2.0) only use disease gene membership information and are not designed to take advantage of disease gene perturbation values. The next few sections describe in detail a set of pairwise similarity metrics — and their corresponding connectivity scores — that have been proposed to address these various limitations and improve the calculation of drug-disease associations [20–23] (**Table 6**).

## Connection Strength Score ($CSS$)

Zhang and Gant proposed a connectivity score called the Connection Strength Score ($CSS$) [23]. Similar to other scores, $CSS$ is formulated to keep each gene's contribution proportional to its level of differential expression. In addition, the goals of this new score are i) to include the perturbation of all the genes in characterizing the effect of a drug (or disease) and ii) to treat gene perturbation in either direction (up or down) equally and together. In this subsection, we present the formulation of $CSS$ using our new, consistent notation (**Table 6**).

These motivations led the authors to propose a new scheme for ranking drug genes. In this scheme, all genes, irrespective of the direction of perturbation, are first ranked in descending order based on the absolute value of their drug gene scores. We represent this operation using the function $r_{drg}^{abs}()$ that takes one or more genes as input and returns their absolute-value-based ranks in the drug profile. Positive or negative signs are then added back to the rank of each up- or down-regulated gene, respectively, to get signed ranks, denoted as $r_{drg}^{abs}() \times sgn_{drg}()$. Similarly, the signed ranks for the disease data are obtained as $r_{dis}^{abs}() \times sgn_{dis}()$. Thus, the $i^{th}$ gene in $R$ (or $S$) gets a signed rank of $N - i + 1$ or $-(N - i + 1)$ depending on whether the gene is up- or down-regulated. These signed ranks are used to calculate a similarity metric between an ordered drug profile $R$ and an ordered disease signature $S$. We refer to this metric as the raw connection strength score ($css$).

$$css(\vec{R}, \vec{S}) = \sum_{i=1}^{N_S} [(r_{drg}^{abs}(gs_i) \times sgn_{drg}(gs_i)) \times (r_{dis}^{abs}(gs_i) \times sgn_{dis}(gs_i))] \tag{5.1}$$

The raw score ($css$) is then scaled by the maximum possible score given the number of drug and disease genes ($css_{max}^{o}(N_R, N_S)$) to calculate a connectivity score that is referred to here as the final connection strength score (CSS).

$$CSS = \frac{css(\vec{R}, \vec{S})}{css_{max}^{o}(N_R, N_S)} \tag{5.2}$$

where $css_{max}^{o}(N_R, N_S) = \sum_{i=1}^{N_S}(N_R - i + 1)(N_S - i + 1)$

Here, genes perturbed in the same direction (up or down) by both the drug and the disease make a positive contribution to $CSS$, while the contribution of genes perturbed in different directions will be negative. Consequently, gene signatures with mixed perturbations will result in an overall low $CSS$ with the positive and negative contributions canceling each other.

As proposed by the authors, this scoring scheme can be easily adapted to the case when only an unordered gene set ($S$) is available for the disease.

$$css(\vec{R}, S) = \sum_{i=1}^{N_S} [r_{drg}^{abs}(gs_i) \times sgn_{drg}(gs_i)] \tag{5.3}$$

$$CSS = \frac{css(\vec{R}, S)}{css_{max}^{u}(N_R, N_S)} \tag{5.4}$$

where $css_{max}^{u}(N_R, N_S) = \sum_{i=1}^{N_S}(N_R - i + 1)$

**Summary**

- Connection strength score, $CSS$, ranges from –1 to +1.
- $CSS$ of +1 and –1 indicate the maximum positive and negative connection strengths, respectively, corresponding to the strongest and weakest possible correlation of the disease profile with the treatment instance used in generating $\vec{R}$ (**Fig. 3**).
- Revised notations used in to define $CSS$ are summarized in **Table 6**.

**Table 6. *CSS* Notations**

| Current Notation | Previous Notation | Description |
|---|---|---|
| $CSS$ | $c$ | Connection Strength Score |
| $\vec{R}$ | $R$ | rank-ordered drug list (*i.e.*, reference profile) |
| $S$ | $s$ | unordered disease signature (*i.e.*, disease gene set) |
| $\vec{S}$ | $s$ | ordered disease signature (*i.e.*, disease gene list) |
| $gs_i$ | $g_i$ | $i^{th}$ gene in set $S$ |
| $css(\vec{R},S)$, $css(\vec{R},\vec{S})$ | $C(R,s)$ | raw connection strength score between $\vec{R}$ and $S$ or between $\vec{R}$ and $\vec{S}$ |
| $r_{drg}^{abs}(gs_i) \times sgn_{drg}(gs_i)$ | $R(g_i)$ | the signed position of $gs_i$ in $\vec{R}$ |
| $r_{dis}^{abs}(gs_i) \times sgn_{dis}(gs_i)$ | $S(g_i)$ | the signed position of $gs_i$ in $\vec{S}$ |
| $sgn_{dis}(gs_i)$ | $S(g_i)$ | disease gene's regulation status in disease profile ($S$ or $\vec{S}$); assigned to +1 and –1 for genes with up- and down-regulation status, respectively |
| $N_S$ | $m$ | number of genes in $S$ that appears in $\vec{R}$ (equivalent to number of genes in $S$ since $S \subseteq R$) |

## Similarity Metrics Based on Differential Expression Values

Though $CSS$ uses all genes from the drug and the disease data, it is still rank-based. Hence, another class of metrics has been proposed to explicitly use the differential expression values of genes in calculating drug-disease similarity [20–22]. As these metrics are simple, their definitions in the original studies are only descriptive. Nevertheless, here, we describe them using our notations to easily relate them to all other metrics and scores.

### *Whole and extreme metrics*

The metric $Sum$ is calculated as the difference between the sum of the drug perturbation values of all up-regulated disease genes ($v_{drg}(S^+)$) and the sum of the drug perturbation values of all down-regulated disease genes. Thus, $Sum = sum(v_{drg}(S^+)) - sum(v_{drg}(S^-))$ (6.1). The metric $Cosine$ captures the cosine correlation between the drug and the disease perturbation values across the common set of all genes with both data, *i.e.*, $Cos(v_{dis}(S), v_{drg}(R))$ (6.2). Analogous similarity metrics can be calculated by replacing cosine correlation with Pearson ($Cor$) and Spearman ($Spe$) correlation coefficients. These metrics can also be adapted to just use a fixed number of 'extreme' genes that are most up- or down-regulated by the disease (as depicted in **Fig. 1**): $XCor$ is the extreme Pearson correlation, defined as $Cor(v_{dis}(S_X), v_{drg}(R_X))$ (6.3); $XSpe$ is the extreme Spearman rank correlation, defined as $Spe(\overrightarrow{S_X}, \overrightarrow{R_X})$ (6.4); $XSum$ is the extreme $Sum$, defined as $sum(v_{drg}(S_X^+)) - sum(v_{drg}(S_X^-))$ (6.5); and $XCos$ is the extreme cosine correlation, defined as $Cos(v_{dis}(S_X), v_{drg}(R_X))$ (6.6).

**Summary**

- $Sum$ and $Cosine$ are pairwise similarity metrics that use gene differential expression values of all genes.
- The extreme similarity metric $XSum$ uses the drug differential expression values of genes most perturbed by the disease. $XCor$, $XSpe$, and $XCos$ compare the disease and drug differential expression values of genes most perturbed by the disease.
- $Sum$ and $XSum$ can take any real value with negative values indicating an overall reversal of the disease perturbation by the drug. $XCor$, $XSpe$, and $XCos$ range from –1 to +1 indicating the maximum positive and negative similarity, respectively, between the drug and the disease (**Fig. 3**).

## Expression-weighted Cosine Similarity ($EWCos$)

The correlation metrics described above take as input the differential expression values of genes from the drug and the disease, which are calculated by comparing drug/disease samples to appropriate controls. These differential expression values, however, do not preserve information about the basal expression levels of the genes. Further, when sample sizes are small, biological and technical noise could result in genes with overall low expression levels ending up with high differential expression levels just by chance. The Expression-Weighted Cosine similarity metric ($EWCos$) was introduced to mitigate the effect of lowly expressed genes [22]. This metric is described next using notations similar to the ones used in the original study (**Table 7**).

First, using data in CMap 2.0, a weight $w_{ij}$ is computed for each gene $gr_i$. for each drug instance $j$ using a logistic sigmoidal function. This function, defined as follows, assigns genes that are lowly or highly expressed with weights close to zero or one, respectively.

$$w_{ij} = \frac{1}{1 + e^{-\alpha(x_{ij} - k\bar{x})}}$$

where $x_{ij}$ is the raw expression value of $gr_i$ in drug instance $j$. $\bar{x}$ is the average of all the raw expression values of the genes in the CMap 2.0 database. $\alpha \in [0,6]$ and $k \in [0,1.5]$ are parameters to be optimized.

The weights ($w_{ij}$) for all the genes across all the drugs are gathered into a matrix $W$. From the drug perturbation data, there is also a matrix $logFC$. Each cell in this matrix $logFC_{ij}$ contains the log-fold-change of gene $gr_i$ in drug instance $j$. Hence, to calibrate log-fold-changes using expression-based weights, these two matrices, $W$ and $logFC$ are combined by element-wise multiplication (Hadamard product) to obtain a matrix of expression-weighted log-fold-changes, $EWlogFC$.

$$EWlogFC = W \circ logFC$$

Finally, given a query disease profile ($S$), quantifying its similarity to each column in the $EWlogFC$ matrix will reveal the association between that disease and each drug in the database. The disease-drug similarity used here is $EWCos$, defined as the cosine similarity between the vector of $logFC$ values of the disease (e.g., $v_{dis}(S)$) and the column in the $EWlogFC$ matrix corresponding to that specific drug $j$.

$$EWCos_j = Cos(v_{dis}(S), EWlogFC_j) \qquad (6.7)$$

As large-scale drug-disease gold-standards are lacking, the $\alpha$ and $k$ parameters in the weight function above are optimized to maximize the ability of $EWCos$ to match replicate instances of the same drug (see [22] for more details).

**Summary**

- Expression-Weighted Cosine metric, $EWCos$, ranges from –1 to +1 with negative values corresponding to drug reversal of disease signature (**Fig. 3**).
- The weighting based on basal expression reduces the contribution of lowly-expressed genes to the drug-disease similarity measure.
- Revised notations used to define $EWCos$ are presented in **Table 7**.

**Table 7. $EWCos$ Notations**

| Current Notation | Previous Notation | Description |
| --- | --- | --- |
| $i, j$ | — | indices of genes and drug instances in CMap 2.0; $i = 1, 2, \ldots, N_S$, $j = 1, 2, \ldots, N_D$ |
| $w_{ij}$ | $w_i$ | weight calculated using the logistic sigmoidal function given specifically to each gene $gr_i$ for drug instance $j$ in the CMap 2.0 |
| $x_{ij}$ | $x_i$ | raw expression value of each gene $gr_i$ for drug instance $j$ in CMap 2.0 |

# Ensemble of Multiple Drug Repositioning Approaches ($EMUDRA$)

Given the range of similarity metrics and connectivity scores that have been developed over the years, going forward, a particularly appealing approach is to integrate multiple metrics to build on each other's strengths and buffer for the weaknesses. With such a goal in mind, Zhou and colleagues proposed $EMUDRA$, an Ensemble of Multiple Drug Repositioning Approaches. $EMUDRA$ combines the similarity metric they developed — $EWCos$ (described above) — with three other pairwise metrics previously shown to perform well [20,21] — $Cosine$, $XCor$ and $XSpe$ — into an integrated prediction model [22]. In this section, we present the formulation of $EMUDRA$ using notations that are identical to the ones used in the original paper.

## $EMUDRA$ Calculation

$EMUDRA$ combines $EWCos$, $Cosine$, $XSpe$, and $XCor$ by first standardizing each score (*i.e.,* subtracting mean and dividing by standard deviation) and summing the resulting $z$-scores of the four metrics to get a final prediction score.

To check if the standardization can be applied directly for each similarity metric, the authors examined if the similarities of all the drugs to a given disease signature follow a normal distribution. For random queries, their similarities to all the drugs were observed to closely follow a normal distribution. On the other hand, for a real disease query, the similarities were observed to be nearly normal except for a long tail corresponding to the few drug instances in the database that effectively reverse the disease signature. Consequently, the similarity scores for a real query signature are standardized using trimmed (winsorized) mean and standard deviation as follows. Let $l_i$ be a list of similarity scores of all the drugs in the database for a given query disease signature, where the index i refers to one of the four different similarity metrics ($i = 1, 2, 3, 4$). Let $Q1, Q3$ be the first and third quartiles of $l_i$, respectively, and $IQR$ be the interquartile range, $(Q3 - Q1)$. The thresholds $[Q1 - (1.5 \times IQR), Q3 + (1.5 \times IQR)]$ are then used to identify the outliers in $l_i$. Let $l'_i$ be a new list created by excluding the outliers in $l_i$. This trimmed list is used to calculate the mean $\mu(l'_i)$ and standard deviation $\sigma(l'_i)$, which are then used to convert the values in $l_i$ to $z$-scores $z_i$.

$$z_i = \frac{l_i - \mu(l'_i)}{\sigma(l'_i)}$$

After applying this winsorized standardization procedure on the scores from all four methods, the final $EMUDRA$ score is calculated as follows:

$$EMUDRA = \sum_i z_i \tag{7.1}$$

**Summary**

- $EMUDRA$ score can be any real number $[-\infty, +\infty]$.
- Large negative scores indicate drugs that invariably have low scores across all four metrics, signifying drug reversal of the disease signature.
- The notations used to define $EMUDRA$ are identical to those in the original paper.

# Discussion

Connectivity-based drug repurposing is a stellar example of the power of thoughtfully combining computational techniques, experimental design, and high-throughput –omics data. Over the past 15 years, this approach has delivered biomedical insights and therapeutic leads for a variety of diseases including COVID-19 [31,32]. During this time, the available data has seen tremendous growth, for e.g., from the thousands of drug profiles in CMAP 1.0 [7] to > 1 million profiles in LINCS [19]. This growth in data is paralleled by the development of several newer connectivity mapping methods for comparing drug and disease gene signatures as effectively as possible.

As is expected, these methods have been built upon each other over time towards addressing previous limitations, leveraging larger amounts of data, and achieving better performance in prioritizing repurposed drug candidates for diseases. Hence, all these methods share a number of core conceptual and analytical ideas and use similar statistical techniques and quantities. Unfortunately, the original studies that published these methods and the other studies that reused, reviewed, or compared the *quantitative details* of different methods have used inconsistent notations and naming systems to refer to previous methods and their mathematical details. Such variation is a considerable impediment to: a) cogent, detailed understanding of current methods, b) their transparent benchmarking and evaluation, and c) the development of new methods that continue to build on existing ideas.

In this article, we present the most comprehensive and detailed description of all connectivity scores and their relationships. This description is grounded on a consistent and all-inclusive system of notations and definitions for all the ideas and quantities involved. To avoid any confusion, we have also clearly tabulated how any new notation that we develop here corresponds to the notations used in the original studies. As can be seen in the descriptions above and the discussion below, this unified system has enabled us to unambiguously refer to methodological details, make clear connections between methods and studies, and discuss their properties.

In the rest of the discussion, we examine all the connectivity scores in terms of their underlying drug-disease similarity metrics that reveal facets of their biological and practical relevance. Next, we outline the status of current efforts to benchmark similarity metrics and connectivity scores. Finally, we present a forward-looking discussion of recent developments and immediate needs in the broader area of computational drug repurposing, along with how our work fits into this big picture.

## Connectivity scores through the lens of disease-drug similarity metrics

The first step in connectivity mapping is the quantification of the association between a single disease and a single drug based on their gene perturbation profile. Connectivity scores differ from each other in their choice of specific similarity metrics and how they are combined, normalized, and background-corrected. Among these aspects, the choice of similarity metric significantly influences the nature of the connectivity score. **Table 8** shows the list of all similarity metrics used in connectivity mapping to quantify disease-drug associations.

**Table 8. Disease-Drug Similarity Measures**

This table shows the disease- and drug-specific information required for the calculation of the nine similarity metrics: $ES$, $css$, $Sum$, $XSum$, $Cosine$, $XCos$, $XCor$, $XSpe$, and $EWCos$, their symmetry, and their associated connectivity scores. The classic enrichment score $ES$, which is based on the signed KS statistic [16], can take two forms: unweighted (uw) and weighted (w). The unweighted form — as used by the connectivity scores $CS$ [7] and $RGES$ [18] — takes as input the set of most perturbed genes from the disease ($S_X$) and the rank ordering of genes in the drug profile ($\vec{R}$). The weighted form — as used by scores $WCS$, $NCS$, and $\tau$ [19] — takes an additional input of drug gene perturbation values that are used to weight the genes at each position in the ranked profile. Another class of pairwise similarity metrics uses rank and/or differential expression values from both the disease and the drug to calculate disease-drug association. For instance, the connection strength score $css$ [23] uses ranks and perturbation directions of all genes (not just the most perturbed) from the disease ($r_{dis}(S)$, $sgn_{dis}(S)$) and the drug ($r_{drg}(R)$, $sgn_{drg}(R)$). $css$ can be adapted to the case when only unordered disease gene sets are available ($S_X^+$, $S_X^-$). Metrics such as Sum use membership information from the disease ($S^+$, $S^-$) with the level of differential expression of genes from the drug ($v_{drg}(R)$) [20,21]. $Cosine$ and $EWCos$ use the perturbation values of genes from both the disease ($v_{dis}(S)$) and the drug ($v_{drg}(R)$) [22]. $EWCos$ differs from $Cosine$ in how the former mitigates the effect of noisy differential expression signals from genes with low expression [22]. The so-called 'extreme' metrics — $XSum$, $XCos$, $XCor$, and $XSpe$ — are equivalent to their parent versions except that the inputs are restricted to the most perturbed genes [20,21].

| Similarity Metric | Disease Information | Drug Information | Symmetric? | Connectivity Score(s) |
|---|---|---|---|---|
| $ES$ (uw) [16] | $S_X$ | $\vec{R}$ | No | $CS$ [7], $RGES$ [18] |
| $ES$ (w) | $S_X$ | $\vec{R}, v_{drg}(R)$ | No | $WCS$, $NCS$, $\tau$ [19] |
| $css$ (o) [23] | $r_{dis}(S), sgn_{dis}(S)$ | $r_{drg}(R), sgn_{drg}(R)$ | Yes | $CSS$ [23] |
| $css$ (u) | $S_X^+, S_X^-$ | $r_{drg}(R), sgn_{drg}(R)$ | No | $CSS$ |
| $Sum$ [20,21] | $S^+, S^-$ | $v_{drg}(R)$ | No | - |
| $XSum$ [20,21] | $S_X^+, S_X^-$ | $v_{drg}(R_X)$ | No | - |
| $Cosine$ [22] | $v_{dis}(S)$ | $v_{drg}(R)$ | Yes | $EMUDRA$ [22] |
| $XCos$ [20,21] | $v_{dis}(S_X)$ | $v_{drg}(R_X)$ | Yes | - |
| $XCor$ [20,21] | $v_{dis}(S_X)$ | $v_{drg}(R_X)$ | Yes | $EMUDRA$ |
| $XSpe$ [20,21] | $\vec{S_X}$ | $\vec{R_X}$ | Yes | $EMUDRA$ |
| $EWCos$ [22] | $v_{dis}(S)$ | $v_{drg}(R)$ | Yes | $EMUDRA$ |

| Drug profile | Disease | | Outcome | ES_up | ES_down | Pairwise similarity measures |
|---|---|---|---|---|---|---|
| | Up | Down | | | | |
| + | | | | − | + | − |
| − | | | | | | |
| + | | | | close to zero | + | − |
| − | | | | | | |
| + | | | | − | close to zero | − |
| − | | | | | | |
| + | | | | + | + | close to zero |
| − | | | | | | |
| + | | | | − | − | close to zero |
| − | | | | | | |
| + | | | | + | close to zero | + |
| − | | | | | | |
| + | | | | close to zero | − | + |
| − | | | | | | |
| + | | | | + | − | + |
| − | | | | | | |

**Figure 3. Similarity metrics and the drug reversal phenotype.** The figure shows expected signs of the disease-drug similarity metrics (last three columns) for all eight scenarios (rows) of overlap between the drug and disease signatures (depicted in first and second columns), each leading to drug reversal outcomes of different strengths and directions (depicted in the "Outcome" column). Specifically, these scenarios correspond to combinations of up- and down-regulated disease genes ($S$) and their relative position in the drug profile ($\vec{R}$). The top three scenarios (coded in blue) correspond to favorable outcomes of the drug fully or partially reversing the disease gene signature. The bottom three scenarios (coded in red) correspond to unfavorable outcomes of the drug not reversing the disease gene signature. The middle two scenarios (coded in grey) indicate neutral outcomes. $ES_{up}$ and $ES_{down}$: enrichment scores of up- and down-regulated disease genes, respectively; Pairwise similarity metrics: collectively refers to $css$, $Sum$, $XSum$, $Cosine$, $XCos$, $XCor$, and $EWCos$.

The differences between similarity metrics (**Table 8**; **Fig. 3**) have a number of biological and practical implications:

**Nature of drug reversal**: The enrichment score $ES$ is typically calculated separately for the genes up- and down-regulated in the disease. Negative values of $ES_{up}$ and positive values of $ES_{down}$ indicate the desired reversal of the disease signature by the drug under consideration (**Fig. 3**). The connectivity scores $CS$ (CMap 1.0) and $WCS/NCS/\tau$ (CMap 2.0) use difference between these two values (*i.e.*, $ES_{up} - ES_{down}$) in further calculation if their signs differ, and zero otherwise. This way, $CS$ and $WCS/NCS/\tau$ take negative values when both the up- and down-regulated disease gene sets are reversed by the drug, positive values when both sets are not reversed, and zero when the reversal is mixed. Though these properties seem biologically meaningful, the RGES study [18] noticed that, when several drugs for a particular disease are considered together, their $CS$ scores do not correlate with their efficacies ($IC_{50}$ values). To satisfy this expected correlation, $RGES$ compares the absolute values of $ES_{up}$ and $ES_{down}$ and takes negative values when $|ES_{up}| < |ES_{down}|$, positive values when $|ES_{up}| > |ES_{down}|$, and zero when they are equal to each other. Calculated this way, $RGES$ values of drugs turn out to be inversely correlated with their efficacies while the sign of $RGES$ alone is not informative about drug-disease reversal anymore. Finally, the pairwise similarity metrics – $css$, $Sum$, $XSum$, $Cosine$, $XCos$, $XCor$, and $EWCos$ — and the connectivity scores that incorporate them — $CSS$ and $EMUDRA$ — have a simple correspondence to the reversal phenotype: the range from negative to positive scores correspond to the range from strong reversal to strong similarity.

**Amount of input information**: $ES$, $Sum$, $XSum$, and the unordered $css$ only require the list of most up- and down-regulated genes from the disease. They are designed for the scenario in which the full gene expression data is available for the drug perturbation and only limited data, typically just gene membership information, is available for the disease perturbation. Hence, these metrics are the easiest to apply for drug repurposing because: a) the CMap and LINCS resources that are typically used as the source of drug perturbation data are available in full, and b) gene membership information can be unearthed even from supplementary tables of disease gene expression studies. None of the other metrics can be used in such cases. Metrics such as $Cosine$, $XCos$, $XCor$, and $EWCos$ use the most amount of information from the disease and the drug, which necessitates access to the full differential expression profiles.

**Choices of threshold parameters**: ES and the extreme similarity metrics require a choice of threshold used to determine the genes most perturbed by the drug or the disease. This choice could be based on the level of significance (*e.g., $p-value < 0.01$*), fold-change (*e.g., $|log_2(FoldChange)| > 1$*), and/or just rank (*e.g.,* top and bottom 100 genes). In any scenario, this choice is likely to significantly influence the performance of each metric in prioritizing real disease-drug associations [21,27].

**Symmetry**: By virtue of being symmetric, the metrics $css$, $Cosine$, $XCos$, $XCor$, $XSpe$, and $EWCos$ can be directly applied to not just disease-drug associations but also to quantify disease-disease (*e.g.,* [33]) and drug-drug relationships (*e.g.,* [12,13]). The other metrics too can be used for these purposes by, for instance, averaging the two asymmetric quantities $similarity(drug_1, drug_2)$ and $similarity(drug_2, drug_1)$ (*e.g.,* [21]).

## *Benchmarking similarity metrics and connectivity scores*

All the similarity metrics and connectivity scores described here have not been systematically benchmarked and compared on a large-scale. One of the biggest challenges in doing so is the lack of a gold standard drug-indication set that spans the drugs in the LINCS collection over a variety of diseases. Therefore, studies often use pairs of drugs that share ATC codes or the same drug profiled independently in CMap and LINCS for benchmarking the methods and follow the evaluations with the analysis of a few individual disease datasets with known associated drugs.

Nevertheless, these comparative studies have shown that simple metrics like $XSum$ and $XCos$ outperform $ES$-based methods [20–22]. As expected, ensemble approaches such as $EMUDRA$ that combine multiple metrics have been shown to perform better than any single metric [22]. In this study, connectivity scores based on $ES$ and $css$ performed poorly. Another study found that the performance of $ES$ and $css$ relative to each other depends on the number of genes in the disease signature [27]. Being rank-based, these metrics could suffer from the contribution genes that are highly ranked but not substantially differentially expressed.

## *Looking forward*

With continued growth in computational and experimental technologies, connectivity mapping remains integral to a number of newer avenues for therapeutic design and applications. Connectivity-based methods have been valuable for comparing drugs to each other based on the similarity of their expression signatures [12,13]. Integrating these drug-drug similarities with drug-disease reversal relationships, both calculated using connectivity scores, has been shown to be powerful in prioritizing synergistic drug combinations [34]. Connectivity score methods are powerful in characterizing the relationship between diseases and the overlap with drugs at the level of perturbed pathways instead of genes [33]. Connectivity has also been adapted for use on drug profiles that are not based on gene expression; for example, drug profiles derived by integrating known chemical-protein associations from several databases [35]. Other new applications are taking connectivity mapping methods into personalized medicine [36,37]. For instance, network-based methods have been used to personalize drug repurposing using patient-specific gene expression data and known gene interactions [36]. Personalized drug repurposing has also been performed using pathway-level (instead of gene-level) comparisons between diseases and drugs [37]. Given the wide range of data types that can be exploited for drug repurposing, the strength now lies in consolidating connectivity mapping methods with other methods and resources to exploiting the variety of signals [38]. Adopting supervised machine learning techniques is going to be key in building the massive frameworks needed for integrative drug repurposing [39].

While the development of new methods is exciting, making them practically useful to the biomedical research community at large requires a concomitant development of data and computing infrastructure. New approaches are needed to increase the scope of resources such as LINCS by computationally increasing data coverage to more cell types [40] and more genes across the human genome [41]. We also need newer flexible software tools that can adapt to multiple types of disease gene expression and drug response database schemas [42], as well as software packages that can house multiple computational methods for drug repurposing [43]. These new methods and packages need to be interfaced with continually curated gold-standards of repurposed drugs for systematic benchmarking methods [1].

Also essential to this growing infrastructure are living surveys of methods and databases [6] as well as unified definition and notation systems like the one presented in this article. The scheme developed here will improve the consistency of future methods with existing ones and help clearly establish the provenance of analytical ideas.

# Conclusion

In this article, we have reconciled several key formulations of drug-disease connectivity scores by defining them and their constituent similarity metrics using consistent notation and terminology. Our coherent definition of connectivity scores and their relationships will allow researchers to better understand the current state-of-the-art and to transparently develop and compare new methods in the context of existing ones. To foster long-term adoption and potential collaborations, this article will be hosted in a GitHub repository (https://github.com/JRaviLab/connectivity_scores) that can be edited by the research community to include new methods for connectivity score calculation. The document has been written using RMarkdown [44,45] and distill [46], and rendered as a living document at https://jravilab.github.io/connectivity_scores.

# Funding

This work was primarily supported by US National Institutes of Health (NIH) grants R35 GM128765 to A.K., MSU Diversity Research Network Launch Awards Program to J.R., MSU College of Natural Science Scholarships to K.S., and, in part, by MSU start-up funds to A.K. and J.R.

# Acknowledgments

We are grateful to members of the Ravi and Krishnan labs for feedback on the manuscript.